\documentstyle[twoside]{article}

\catcode`\@=11
\long\def\@makefntext#1{
\protect\noindent \hbox to 3.2pt {\hskip-.9pt  
$^{{\eightrm\@thefnmark}}$\hfil}#1\hfill}		

\def\@makefnmark{\hbox to 0pt{$^{\@thefnmark}$\hss}}	
	
\def\ps@myheadings{\let\@mkboth\@gobbletwo
\def\@oddhead{\hbox{}
\rightmark\hfil\eightrm\thepage}   
\def\@oddfoot{}\def\@evenhead{\eightrm\thepage\hfil
\leftmark\hbox{}}\def\@evenfoot{}
\def\sectionmark##1{}\def\subsectionmark##1{}}

\oddsidemargin=\evensidemargin
\newcounter{sectionc}\newcounter{subsectionc}\newcounter{subsubsectionc}
\renewcommand{\section}[1] {\vspace{12pt}\addtocounter{sectionc}{1} 
\setcounter{subsectionc}{0}\setcounter{subsubsectionc}{0}\noindent 
	{\tenbf\thesectionc. #1}\par\vspace{5pt}}
\renewcommand{\subsection}[1] {\vspace{12pt}\addtocounter{subsectionc}{1} 
	\setcounter{subsubsectionc}{0}\noindent 
	{\bf\thesectionc.\thesubsectionc. {\kern1pt \bfit #1}}\par\vspace{5pt}}
\renewcommand{\subsubsection}[1] {\vspace{12pt}\addtocounter{subsubsectionc}{1}
	\noindent{\tenrm\thesectionc.\thesubsectionc.\thesubsubsectionc.
	{\kern1pt \tenit #1}}\par\vspace{5pt}}

\newcommand{\textlineskip}{\baselineskip=13pt}
\newcommand{\smalllineskip}{\baselineskip=10pt}

\def\eightcirc{
\begin{picture}(0,0)
\put(4.4,1.8){\circle{6.5}}
\end{picture}}
\def\eightcopyright{\eightcirc\kern2.7pt\hbox{\eightrm c}} 

\newcommand{\copyrightheading}[1]
	{\vspace*{-2.5cm}\smalllineskip{\flushleft
      {\footnotesize Quart. J. Math. Oxford (2), 6 (1955), 121-127  #1}\\
      {\footnotesize \LaTeX $~$ by H.C. Rosu (1999); physics/9908019 #1}\\
	 }}

\renewenvironment{thebibliography}[1]
	{\frenchspacing
	 \ninerm\baselineskip=11pt
	 \begin{list}{\arabic{enumi}.}
        {\usecounter{enumi}\setlength{\parsep}{0pt}     
	 \setlength{\leftmargin 12.7pt}{\rightmargin 0pt} 
         \setlength{\itemsep}{0pt} \settowidth
	{\labelwidth}{#1.}\sloppy}}{\end{list}}

\newcounter{itemlistc}
\newcounter{romanlistc}
\newcounter{alphlistc}
\newcounter{arabiclistc}

\def\@citex[#1]#2{\if@filesw\immediate\write\@auxout
	{\string\citation{#2}}\fi
\def\@citea{}\@cite{\@for\@citeb:=#2\do
	{\@citea\def\@citea{,}\@ifundefined
	{b@\@citeb}{{\bf ?}\@warning
	{Citation `\@citeb' on page \thepage \space undefined}}
	{\csname b@\@citeb\endcsname}}}{#1}}

\newif\if@cghi
\def\cite{\@cghitrue\@ifnextchar [{\@tempswatrue
	\@citex}{\@tempswafalse\@citex[]}}
\def\citelow{\@cghifalse\@ifnextchar [{\@tempswatrue
	\@citex}{\@tempswafalse\@citex[]}}
\def\@cite#1#2{{$\null^{#1}$\if@tempswa\typeout
	{IJCGA warning: optional citation argument 
	ignored: `#2'} \fi}}

\def\@refcitex[#1]#2{\if@filesw\immediate\write\@auxout
	{\string\citation{#2}}\fi
\def\@citea{}\@refcite{\@for\@citeb:=#2\do
	{\@citea\def\@citea{, }\@ifundefined
	{b@\@citeb}{{\bf ?}\@warning
	{Citation `\@citeb' on page \thepage \space undefined}}
	\hbox{\csname b@\@citeb\endcsname}}}{#1}}

\def\@refcite#1#2{{#1\if@tempswa\typeout
        {IJCGA warning: optional citation argument
	ignored: `#2'} \fi}}

\def\refcite{\@ifnextchar[{\@tempswatrue
	\@refcitex}{\@tempswafalse\@refcitex[]}}

\def\pmb#1{\setbox0=\hbox{#1}
	\kern-.025em\copy0\kern-\wd0
	\kern.05em\copy0\kern-\wd0
	\kern-.025em\raise.0433em\box0}

\def\fnt#1#2{\footnotetext{\kern-.3em
	{$^{\mbox{\scriptsize #1}}$}{#2}}}

\def\runninghead#1#2{\pagestyle{myheadings}
\markboth{{\protect\footnotesize\it{\quad #1}}\hfill}
{\hfill{\protect\footnotesize\it{#2\quad}}}}
\font\tenrm=cmr10
\font\tenit=cmti10 
\font\tenbf=cmbx10
\font\bfit=cmbxti10 at 10pt
\font\ninerm=cmr9

\font\eightrm=cmr8

\def\qed{\hbox{${\vcenter{\vbox{			
   \hrule height 0.4pt\hbox{\vrule width 0.4pt height 6pt
   \kern5pt\vrule width 0.4pt}\hrule height 0.4pt}}}$}}

\begin{document}

\pagestyle{empty}

\begin{center}
{\it Los Alamos Electronic Archives: physics/9908019}
\end{center}

\bigskip

\begin{center}
{\bf MAKING OLD SEMINAL RESULTS WORLD-WIDE AVAILABLE !}
\end{center}

$\;$\\
$\;$\\
$\;$\\
$\;$\\
$\;$\\
$\;$\\

\begin{center}

{\bf FORWARD}

\end{center}

\bigskip

\noindent
The seminal paper of Crum, published in 1955, is now a standard reference 
in nonlinear science and supersymmetric quantum mechanics. It introduces the
Crum transformations, a cornerstone of integrability and a beautiful 
generalization of Darboux transformations.

\bigskip

\noindent
Since I am sure that many people would like to study carefully this 
masterpiece I offer here a 
LaTex version of the paper. The purpose is to prevent all sorts of rediscoveries 
and promote real progress. I did very minor
changes with respect to the old published version. The most important was to
put the list of references at the end and not as footnotes. Crum's paper has 7
points. The first point is the statement of Crum's theorem, i.e., the 
possibility to write the solutions of a tower of so-called associated
Sturm-Liouville (SL) systems (all of them  Dirichlet from the point of view of 
boundary conditions) as a quotient of Wronskian determinants.
The second point refers to the first associated SL system, dealing in fact with
the SL Darboux transformations. Points 3 and 4 are a detailed study 
of the higher order associated SL systems (SL supersymmetric partners).
Point 5 contains four noted applications. The corollary of Crum's theorem is at 
point 6. Finally, point 7 states the possibility to build a regular SL system 
with any finite set of real numbers as eigenvalues, starting from a given
associated SL system, a remarkable general result.

\bigskip
\bigskip

\hfill ${\cal H}$ ${\cal C}$ ${\cal R}$

\bigskip

\hfill 8. 9. 1999


\newpage

\runninghead{Crum
$\ldots$} {Crum
$\ldots$}


\normalsize\textlineskip
\thispagestyle{empty}
\setcounter{page}{1}

\copyrightheading{}                     

\vspace*{0.88truein}

\centerline{\bf ASSOCIATED STURM-LIOUVILLE SYSTEMS}
\vspace*{0.035truein}
\vspace*{0.37truein}
\centerline{\footnotesize  by  M.M. CRUM ({\em Oxford})}
\vspace*{0.015truein}
\centerline{\footnotesize [Received 7 September 1954] }
\vspace*{0.225truein}



\textlineskip                  
\vspace*{12pt}                 

\vspace*{1pt}\textlineskip	
\vspace*{-0.5pt}
\noindent


\noindent




\noindent

{\bf 1}. Let the regular Sturm-Liouville system
$$
\left\{\begin{array}{lll}
y^{''}+[\lambda -q(x)]y=0 & \hspace{0.5cm} (0<x<1)~, &  \hspace{4.5cm} (A)\\
y^{'}(0)=h^{(0)}y(0)~,    & \hspace{0.5cm}
y^{'}(1)=h^{(1)}y(1) & \hspace{4.5cm} (B)
\end{array} \right.
$$
have eigenvalues $\lambda _{0}<\lambda _{1} < \lambda _{2}$, etc, and
eigenfunctions $\phi _{s}$ corresponding to $\lambda _{s}$. Let  $q(x)$
be repeatedly differentiable in (0,1); then the $\phi _{s}$ also are
repeatedly differentiable; let $W_{ns}$ be the Wronskian of the $n+1$
functions $\phi _{0}$, $\phi _{1}$,..., $\phi _{n-1}$, $\phi _{s}$ and
let $W_{n}$ be the Wronskian of the $n$ functions
$\phi _{0}$, $\phi _{1}$,..., $\phi _{n-1}$. Then, if $n\geq 1$ and
$$
\phi _{ns}=W_{ns}/W_{n}~,
$$
the functions $\phi _{ns}~ (s\geq n)$ are the eigenfunctions, with
eigenvalues $\lambda _{s}$, of the system
$$
\left\{\begin{array}{lll}
y^{''}+[\lambda -q_{n}(x)]y=0 & \hspace{0.5cm} (0<x<1)~, & \hspace{4.5cm}
(A_{n})\\
\lim _{x\rightarrow 0}y(x)=0~,    & \hspace{0.5cm}
\lim _{x\rightarrow 1}y(x)=0 & \hspace{4.5cm} (B_{n})
\end{array} \right.
$$
where
$$
q_{n}(x)=q(x)-2\frac{d^2}{dx^2}\log W_{n}~.
\eqno(C_{n})
$$
For $n=1$, the system $({\rm A_{n}, B_{n}})$ is regular; but, for $n>1$,
$$
q_{n}(x)\approx
\left\{\begin{array}{ll}
n(n-1)x^{-2} & \hspace{0.5cm} (x\rightarrow 0)~, \\
n(n-1)(1-x)^{-2}   & \hspace{0.5cm}
(x\rightarrow 1)
\end{array} \right.
$$
Inside (0,1), $W_{n}$ is non-zero and $q_{n}$ is continuous. For $s<n$,
$\phi _{ns}\equiv 0$; for $s>n$, $\phi _{ns}$ has exactly $s-n$ zeros inside
(0,1). The family $\phi _{ns}~ (s\geq n)$ is $L^2$-closed and complete over
(0,1).

The system $({\rm A_{n},B_{n}})$ may be called the `nth system associated
with the system $({\rm A,B})$'. In this note the above statements are
established, and examples are given of systems associated with non-regular
Sturm-Liouville systems.

If $q(x)$ is continuous but not differentiable, the $\phi _{s}$ are
differentiable twice only, and the Wronskians do not exist; however, when
the Wronskians $W_{ns}$, $W_{n}$ exist, they are equal to the modified
Wronskians $W_{ns}^{*}$, $W_{n}^{*}$ obtained by replacing $\phi _{s}^{(2k)}$
by $(-\lambda _{s})^{k}\phi _{s}$, and
$\phi _{s}^{(2k+1)}$
by $(-\lambda _{s})^{k}\phi _{s}^{'}$; the $W_{n}^{*}$ are at least twice
differentiable, and the statements above are true for
non-differentiable continuous $q$ provided that the $W$ are replaced by
$W^{*}$.

\bigskip
\newpage
{\bf 2}. \underline{The case $n=1$}

\bigskip

\noindent
We have $W_{1}=\phi _{0}$, of constant sign [\refcite{Ince}]
for $0\leq x\leq 1$; and
$$
\phi_{1s}=\phi _{s}^{'}-\frac{\phi _{0}^{'}}{\phi _{0}}\phi _{s}=
\phi _{s}^{'}-v\phi _{s}, {\rm say},
\eqno(D_1)
$$
where
$$
v^{'}+v^2=q-\lambda _{0}~.
\eqno(E)
$$
Then
$$
\frac{d}{dx}\left( \phi _{0}\phi _{1s}\right)=\phi _{0}\phi _{s}^{''}
-\phi _{0}^{''}\phi _{s}=(\lambda _{0}-\lambda _{s})\phi _{0}\phi _{s}~.
\eqno(F_1)
$$
Since
$$
\phi _{1s}(0)=0=\phi _{1s}(1)~,
\eqno(G)
$$
we have
$$
\phi _{0}\phi _{1s}=(\lambda _{0}-\lambda _{s})\int _{0}^{x}\phi _{0}(\xi)
\phi _{s}(\xi)d\xi=-
(\lambda _{0}-\lambda _{s})\int _{x}^{1}\phi _{0}\phi _{s}d\xi~.
\eqno(G^{'})
$$
Hence
\begin{eqnarray*}
\phi _{1s}^{'} & = (\lambda _{0}-\lambda _{s})\phi _{s}-v\phi _{1s}~,\\
\phi _{1s}^{''} & = (\lambda _{0}-\lambda _{s})\phi _{s}^{'}-v^{'}\phi _{1s}-
v[(\lambda _{0}-\lambda _{s})\phi _{s}-v\phi _{1s}]\\
               &=  (\lambda _{0}-\lambda _{s}-v^{'}+v^2)\phi _{1s}\\
               &= (q_1-\lambda _{s})\phi _{1s}~,\\
\end{eqnarray*}
where
$$
q_1=\lambda _{0}-v^{'}+v^2=q-2v^{'}=q-2\frac{d^2}{dx^2}\left(\log W_1\right)~.
$$
Now from $(D_1)$,
$$
\phi _{1s}/\phi _{0}=\frac{d}{dx}\left(\phi _{s}/\phi _{0}\right)~;
$$
since $\phi _{s}$ has exactly $s$ zeros [\refcite{Ince}] inside (0,1), by
Rolle's theorem, $\phi _{1s}$ has at least $s-1$. But from $(F_1)$ and $(G)$
and Rolle's theorem, $\phi _{1s}$ has at most $s-1$ zeros inside (0,1); hence
it has $s-1$ exactly. it follows [\refcite{Ince}] that the $\phi _{1s}~
(s\geq 1)$ are all the eigenfunctions of the regular system $(A_1,B_1)$. For
$\lambda \neq \lambda _{0}$ the general solution of $(A_1)$ is
$$
X_1=W(\phi _{0},\chi)/W_1~,
$$
where $\chi$ is the general solution of $(A)$. For $\lambda =\lambda _{0}$,
$W(\phi _{0},\chi)$ is constant and one solution of $(A_1)$ is $1/\phi _{0}$;
two independent solutions are
$$
\frac{1}{\phi _{0}}\int _{0}^{x} \phi _{0}^{2}(\xi) d\xi~, \qquad \frac{1}
{\phi _{0}}\int _{x}^{1} \phi _{0}^{2}(\xi) d\xi~.
$$
It is easily verified that the only solutions of $(A_1)$ which satisfies
$(G)$ are the $\phi _{1s}~ (s\geq 1)$.

\bigskip

{\bf 3}. \underline{The case $n > 1$}

\bigskip

\noindent
Applying Jacobi's theorem to the determinant $W_{ns}$, we have, for
$n > 1$,
$$
W_{ns}W_{n-1}=W_{n}\frac{d}{dx}W_{n-1,s}-W_{n-1,s}\frac{d}{dx}W_{n}~,
$$
with a similar relation with $W^{*}$ for $W$. Hence
$$
\phi _{ns}=\frac{W_{ns}}{W_{n}}=\frac{1}{W_{n-1}}\frac{d}{dx}\left(W_{n-1}
\phi _{n-1,s}\right)-\phi _{n-1,s}\frac{1}{W_{n}}\frac{d}{dx}W_{n}
$$
$$
=\phi _{n-1,s}^{'}-v_{n-1}\phi _{n-1,s}=\frac{1}{\phi _{n-1,n-1}}
W(\phi _{n-1,n-1},\phi _{n-1,s})~,
\eqno(D_{n})
$$
where
$$
v_{n}=\phi _{nn}^{'}/\phi _{nn}~, \qquad \qquad
v_{n-1}=W_{n}^{'}/W_{n}-W_{n-1}^{'}/W_{n-1}~.
$$
Hence, by steps similar to those of \S 2, and by induction on $n$,
$$
v_{n}^{'}+v_{n}^{2}=q_{n}-\lambda _{n}~,
\eqno(E_{n})
$$
$$
\frac{d}{dx}\left(\phi _{n-1,n-1}\phi _{ns}\right)=
(\lambda _{n-1}-\lambda _{s})\phi _{n-1,n-1}\phi _{n-1,s}~,
\eqno(F_{n})
$$
$$
\phi _{ns}^{''}=(q_{n}-\lambda _{s})\phi _{ns}~, \qquad
q_{n}=q_{n-1}-2v_{n-1}^{'}~,
$$
$$
q_{n}+2\frac{d}{dx}\left(\frac{W_{n}^{'}}{W_{n}}\right)=
q_{n-1}+2\frac{d}{dx}\left(\frac{W_{n-1}^{'}}{W_{n-1}}\right)=q~.
$$
We now prove by induction on $n$ the following:
$$
\phi _{ns}=C_{ns}\prod _{t=0}^{n-1}(\lambda _{t}-\lambda _{s})x^{n}
[1+O(x^2)]\qquad (C_{ns}\neq 0)~,
\eqno(G_{n})
$$
$$
\phi _{ns}^{'}=nx^{-1}\phi _{ns}[1+O(x^2)]~,
\eqno(H_{n})
$$
$$
v_{n}=nx^{-1}[1+O(x^2)]~,
\eqno(J_{n})
$$
all as $x\rightarrow 0$, with similar relations as $x\rightarrow 1$;
$$
\phi _{ns} ~ {\rm has} ~ s-n ~ {\rm zeros} ~ {\rm inside} ~ (0.1)~.
\eqno(K_{n})
$$
By $(K_{n})$, $\phi _{nn}$, and so also $W_{n+1}$, is non-zero inside (0,1),
so that $q_{n+1}$ and $\phi _{n+1,s}$ are continuous inside (0,1). First,
by $(G)$ and $(G^{'})$, as $x\rightarrow 0$,
$$
\phi _{1s}(x)\sim (\lambda _{0}-\lambda _{s})\phi _{s}(0)x~;
$$
also
$$
\phi _{1s}^{''}(0) =(q_1-\lambda _{s})\phi _{1s}(0)=0~,
$$
which together imply $(G_1)$; $(H_{1})$ follows from $(G_1)$ and $(F_1)$,
together with
$$
\phi _{s}=\phi _{s}(0)[1+h^{(0)}x+O(x^2)]~;
$$
and $(J_1)$ is a case of $(H_1)$. It remains to deduce $(G_{n+1})$ to
$(K_{n+1})$ from $(G_{n})$ to $(K_{n})$. First, by $(D_{n+1})$, $(H_{n})$,
$(J_{n})$,
$$
\phi _{n+1,s}=\phi _{ns}\big[\frac{n}{x}+O(x)-\frac{n}{x}+O(x)\big]= o(1)
\quad (x\rightarrow 0)~.
$$
Hence
$$
\phi _{nn}\phi _{n+1,s}=(\lambda _{n}-\lambda _{s})\int _{0}^{x}\phi _{nn}
\phi _{ns}d\xi~,
$$
whence we have $(G_{n+1})$ with
$$
C_{n+1,s}=C_{ns}/(2n+1)\neq 0~.
$$
By differentiating this last we obtain $(H_{n+1})$, of which $(J_{n+1})$ is a
special case.

From $(D_{n+1})$ and $(K_{n})$, $\phi _{n+1,s}$ has at least $s-n-1$ zeros
inside (0,1); from $(F_{n+1})$, $(K_{n})$, $(G_{n})$, it has at most
$s-n-1$ zeros inside (0,1); hence $(K_{n+1})$ is deduced.

Lastly we may prove that, as $x\rightarrow 0$,
$$
q_{n}(x)=n(n-1)x^{-2}+O(1)~,
\eqno(L_{n})
$$
with a similar relation as $x\rightarrow 1$. For, given $(L_{n})$ and
$(J_{n})$,
$$
q_{n+1}=q_{n}-2v^{'}_{n}=2\lambda _{n}+2v_{n}^{2}-q_{n}=O(1)+n(n+1)x^{-2}~,
$$
which is $(L_{n+1})$.

For $\lambda \neq \lambda _{s}~ (s<n)$ the general solution of $(A_{n})$ is
$$
y=\chi _{n}=W(\phi _{0},\phi _{1},...,\phi _{n-1},\chi)/W_{n}~,
$$
where $\chi$ is the general solution of $(A)$. For $\lambda =\lambda _{n-1}$
a solution is
$$
y=\frac{1}{\phi _{n-1,n-1}}W(\phi _{n-1,n-1},\chi _{n-1,n-1})=
\frac{C}{\phi _{n-1,n-1}}=C\frac{W(\phi _{0},\phi _{1},..., \phi _{n-2})}
{W(\phi _{0},\phi _{1},...,\phi _{n-1})}~.
$$
For $\lambda =\lambda _{s}$, $s\leq n-1$, a solution is
$$
y=\psi _{ns}=W_{n}^{(s)}/W_{n}~,
$$
where $W_{n}^{(s)}$ is the Wronskian of the $n-1$ functions
$$
\phi _{t} \quad (0\leq t\leq n-1~;~ t\neq s)~.
$$

\bigskip

{\bf 4}. Since the system ${\rm (A_{n}, B_{n})}$ is not regular for $n>1$, it
remains to prove that the family $\phi _{ns}\quad (s\geq n)$
is $L^{2}$-complete over (0,1); this implies incidentally that
the $\phi _{ns}$ are the only bounded solutions of $({\rm A_{n}})$. Since
$({\rm A _{1}, B_{1}})$ is regular, it is sufficient to verify that the
completeness of the family $\phi _{ns}$ implies that of the
family $\phi _{n+1,s}$.

Let $f(x)$ be of $L^2(0,1)$; then, given $\epsilon > 0$, there exists $g(x)$
such that

(i) $g(x)=0 \qquad (0~<~x~<~\delta ~;~ 1-\delta ~<~x~<~1~; \delta ~>~0)$,

(ii)$g^{'}(x)$ is continuous in (0,1),

(iii) $\int _{0}^{1}|f-g|^2d\xi ~<~ \epsilon$.

\noindent
Then, if
$$
h=g^{'}+v_{n}g~, \qquad \qquad \phi _{nn}h=\frac{d}{dx}
\left(\phi _{nn}g\right)~,
$$
h is of $L^2(0,1)$; also
$$
\int _{0}^{1}h\phi _{nn}d\xi= [g\phi _{nn}]_{0}^{1}=0~,
$$
so that, assuming the completeness of the family $\phi _{ns}$, we have
$$
h=\sum _{s=n+1}^{N}c_{s}\phi _{ns}+\eta~,
$$
where
$$
\int _{0}^{1}|\eta ^2|dx <\epsilon ~.
$$
Now
$$
\phi _{nn}g=\int _{0}^{x}\phi _{nn}hd\xi=\sum _{s=n+1}^{N}c_{s}
\int _{0}^{x}\phi _{nn}\phi _{ns}d\xi +\int _{0}^{x}\phi _{nn}\eta d\xi =
\phi _{nn}\sum _{s=n+1}^{N}C_{s}\phi _{n+1,s}+\phi _{nn}\zeta~,
$$
where $C_{s}=c_{s}(\lambda _{n}-\lambda _{s})^{-1}$, and
$$
\zeta =\frac{1}{\phi _{nn}}\int _{0}^{x}\phi _{nn}\eta d\xi
=-\frac{1}{\phi _{nn}}\int _{x}^{1}\phi _{nn}\eta d\xi~;
$$
since, by $(G_{n})$ and its analogue for $x\rightarrow 1$,
$$
\int _{0}^{x}\phi _{nn}^{2}dx=O(\phi _{nn}^{2})~, \quad
\int _{x}^{1}\phi _{nn}^{2}=O(\phi _{nn}^{2})
$$
when $x\rightarrow 0,~1$, respectively,
we have by Schwartz's inequality
$$
|\zeta ^{2}|<M_{n}\int _{0}^{1}|\eta ^{2}|dx<M_{n}\epsilon~, \quad
\int _{0}^{1}|\zeta ^{2}|dx<M_{n}\epsilon~.
$$
Hence the result.

\bigskip
\newpage

{\bf 5}. \underline{Examples}

\bigskip

(1) If $q(x)=0$, $h^{(0)}=0=h^{(1)}$, then $\lambda _{s}=(2\pi s)^2$,
$\phi _{s}=\cos 2\pi sx$ ($s=0,1,2,...$). Since $v=0$, $q_1=q$ and
$$
\phi _{1s}=\phi _{s}^{'}=2\pi s \sin 2\pi s x \qquad \qquad (s=1,2,...)~.
$$
For $n>1$, $\phi _{ns}$ is obtainable as in Example 3.

(2) If $q(x)=x^2$ and the interval is $(-\infty, \infty)$, (A) is
$y^{''}+(\lambda-x^2)y=0$, with $\phi _{0}=e^{-\frac{1}{2}x^2}$,
$\lambda _{0}=1$. Since $v=x$, $q_{1}=x^2-2$; hence [\refcite{Di}]
$$
\lambda _{s+1}=\lambda _{s}+2~, \qquad \qquad  \phi _{1s}=k_{s}\phi _{s-1}~.
$$
The associated systems are all identical, $\lambda _{s}=2s+1$, and, since
$$
\phi _{0}\phi _{s}=\frac{1}{\lambda _{0}-\lambda _{s}}\frac{d}{dx}\left(
\phi _{0}\phi _{1s}\right)=\frac{k_{s}}{2s}\frac{d}{dx}\left(\phi _{0}
\phi _{s-1}\right)~,
$$
it follows that
$$
\phi _{s}=K_{s}\phi _{0}^{-1}\left(\frac{d}{dx}\right)^{s}\phi _{0}^{2}=
K_{s}e^{\frac{1}{2}x^2}\left(\frac{d}{dx}\right)^{s}e^{-x^2}~.
$$

(3) The Legendre functions [\refcite{T}]
$$
y_{s}=(\sin \theta)^{\frac{1}{2}}P_{s}(\cos \theta) \qquad \qquad
(0<\theta < \pi)
$$
satisfy
$$
y^{''}+\left(\lambda +\frac{1}{4}{\rm cosec}^{2}\theta\right)y=0~,
$$
where
$$
\lambda _{s}=(s+\frac{1}{2})^{2} \qquad \qquad (s=0,1,2,...)~.
$$

Writing $\mu =\cos \theta$, and $W_{(\mu)}$ for the Wronskians with respect
to $\mu$, we get
$$
W_{n}=W(y_{0},y_1,...,y_{n-1})=\left(\frac{d\mu}{d\theta}\right)^{\frac{1}{2}
n(n-1)}W_{(\mu)}(y_0,y_1,...,y_{n-1})
$$
$$
=
\left(\frac{d\mu}{d\theta}\right)^{\frac{1}{2}
n(n-1)}(\sin \theta)^{\frac{1}{2}n}W_{(\mu)}(P_0,P_1,...,P_{n-1})=
A_{n}(\sin \theta)^{\frac{1}{2}n^2}~,
$$
and similarly
$$
W_{ns}=A_{n}(\sin)^{\frac{1}{2}(n+1)^2}\left(\frac{d}{dx}\right)^{n}
P_{s}(\mu)~.
$$
Hence [\refcite{WW}]
$$
\phi _{ns}=(\sin \theta)^{n+\frac{1}{2}}\left(\frac{d}{dx}\right)^{n}
P_{s}(\mu)=(\sin \theta)^{\frac{1}{2}}P_{s}^{(n)}(\mu)~.
$$

(4) For the Hankel system [\refcite{T2}] of order $\nu$
$$
y=\phi _{k}(x)=c_{k}(kx)^{\frac{1}{2}}J_{\nu}(kx)~, \qquad \qquad
\phi _{0}(x)=x^{\nu +\frac{1}{2}}~,
$$
$$
y^{''}+\left(\lambda -\frac{\nu ^2-\frac{1}{4}}{x^2}\right)y=0~, \qquad \qquad
\lambda =k^2~.
$$
Here $v=\phi _{0}^{'}/\phi _{0}=(\nu +\frac{1}{2})/x$, whence
$$
q_{1}=\frac{(\nu+1)^2-\frac{1}{4}}{x^2}
$$
and the first associated system is the Hankel system of order $\nu +1$.

\bigskip

{\bf 6}. As a corollary of the main theorem, if
$$
S(x)=\sum _{0}^{n} c_{s}\phi _{s}(x)~,
$$
then $S(x)$ has at most $n$ zeros in (0,1). This result is due to Kellogg
[\refcite{K}]. For, if $S(x)$ has $k$ zeros, then by Rolle's theorem
$$
S_{1}(x)=\phi _{0}\frac{d}{dx}\left( \phi _{0}^{-1}\sum _{0}^{n}c_{s}
\phi _{s}(x)\right)=\sum _{1}^{n}c_{s}\phi _{1s}
$$
has at least $k-1$ zeros inside (0,1); by induction
$$
S_{m}(x)=\sum _{m}^{n}c_{s}\phi _{ms}
$$
has at least $k-m$ zeros, and $S_{n}(x)=c_{n}\phi _{nn}$ has at least $k-n$;
since $\phi _{nn}$ is non-zero, either $k\leq n$ or $c_{n}=0$; but, if
$c_{n}=0$, then $k\leq n-1\leq n$.

This proof of the corollary depends only on the fact that the Wronskians
$W_{n}$ are non-zero. If $\phi _{s}=e^{\alpha _{s} x}$, where the $\alpha _{s}$
are any distinct real numbers, then the $W_{n}$ are all non-zero, and so
$S(x)$ has at most $n$ real zeros.

\bigskip

{\bf 7}. If ${\rm (A,B)}$ is given, the associated systems
${\rm (A_{n}, B_{n})}$ are uniquely defined; but to a given ${\rm (A_{n},
B_{n})}$ belong an infinity of ${\rm (A,B)}$. For example, given
${\rm (A_1,B_1)}$ we may solve for $v$
$$
\lambda _{0}-v^{'}+v^{2}=q_{1}~,
$$
with any $\lambda _{0}$ such that $\lambda _{0}<\lambda _{1}$; then, if
$$
\phi _{0}=\exp \left(\int _{0}^{x}vd\xi\right),\qquad \qquad
(\lambda _{0}-\lambda _{s})\phi _{0}\phi _{s}=\frac{d}{dx}\left(\phi _{0}
\phi _{1s}\right)~,
$$
it will follow that the $\phi _{s}$ are the eigenfunctions of $({\rm A,B})$
with
$$
q=q_1+2v^{'}~, \qquad \qquad h^{(0)}=v(0)~, \qquad \qquad h^{(1)}=v(1)~.
$$
For example, if
$$
q_{1}=0~, \qquad \qquad \lambda _{s}=(2\pi s)^2~, \qquad \qquad \phi _{1s}=
\sin 2\pi sx~,
$$
we can take
$$
\lambda _{0}=-\rho ^{2}~,\qquad \qquad \phi _{0}={\rm sech} \rho(x-\alpha)~,
\qquad \qquad v=-\rho {\rm tanh} \rho (x-\alpha)~,
$$
$$
q(x)=-2\rho ^2 {\rm sech}^{2} \rho (x-\alpha)~,
$$
$$
\phi _{s}(x)=2\pi s \cos 2\pi s x -\rho {\rm tanh} \rho (x-\alpha)
\sin 2\pi s x~.
$$

Starting from a given ${\rm (A_{n},B_{n})}$ we can similarly construct an
${\rm (A,B)}$ with arbitrary
$\lambda _{0}$, $\lambda _{1}$,..., $\lambda _{n-1}$ (provided only that
$\lambda _{s+1}>\lambda _{s}$). Thus there exists a regular Sturm-Liouville
system with any finite set of real numbers as eigenvalues.

\bigskip
\bigskip

\end{document}